\newcommand{\const}{\mathop{\mathrm{const}}\nolimits}
\newcommand{\tr}{\mathop{\mathrm{Tr}}\nolimits}
\newcommand{\Hom}{\mathop{\mathrm{Hom}}\nolimits}

\def\beq{\begin{equation}}
\def\eeq{\end{equation}}
\def\beqn{\begin{eqnarray}}
\def\eeqn{\end{eqnarray}}

\documentclass[12pt]{article}
\pdfoutput=1
\usepackage{amsmath,amssymb}
\usepackage[matrix,arrow,curve]{xy}
\usepackage{graphicx}

\begin{document}
\begin{titlepage}

\begin{flushright}
ITEP- 35/13
\end{flushright}

\vspace{1cm}

\begin{center}
{  \Large \bf  BPS states in the $\Omega$-background  and torus knots}
\end{center}
\vspace{1mm}

\begin{center}

 {\large
 \bf   K.Bulycheva$\,^{a}$ and A.Gorsky$\,^{a,b}$}

\vspace{3mm}

$^a$
{\it Institute of Theoretical and Experimental Physics,
Moscow 117218, Russia}\\[1mm]
$^b$
{\it Moscow Institute of Physics and Technology,
Dolgoprudny 141700, Russia}
\end{center}

\centerline{\small\tt gorsky@itep.ru }
\centerline{\small\tt bulycheva@itep.ru }

\vspace{1cm}

\begin{center}
{\large\bf Abstract}
\end{center}

We clarify some issues concerning the central charges saturated by the extended objects in the SUSY $U(1)$ $4d$ gauge theory in the $\Omega$-background.
The configuration involving the monopole localized at the domain wall is considered in some details. At the rational ratio
$\frac{\epsilon_1}{\epsilon_2}=\frac{p}{q}$ the trajectory of the monopole provides the torus $(p,q)$ knot in the squashed three-sphere.
Using the relation between the integrable systems of Calogero type  at the rational couplings and the torus knots we
interpret this configuration in terms of the auxiliary $2d$ quiver  theory or $3d$ theory with nontrivial
boundary conditions. This realization can be considered as the AGT-like representation of the torus knot invariants.

\end{titlepage}

\tableofcontents

\section{Introduction}

The BPS states provide the useful laboratory for investigation the dynamics of the stable defects of  different codimensions.
Their classification is governed by the corresponding central charges in the SUSY algebra which in four dimensions involves particles, strings and domain walls.
In the paper \cite{bcgk} (see also \cite{ito,hellerman}) we have classified the central charges in the $\Omega$-deformed $\mathcal N=2$ SYM theory. These involve all types of defects and corresponding central charges. When  tensions of the corresponding objects tend to infinity they provide the corresponding boundary conditions and become non-dynamical \cite{gukwit}. In this paper we clarify a few subtle points from the previous analysis and focus at the particular configuration corresponding to the monopole localized at the domain wall.
This is an example of the line operator at the domain wall  considered in  \cite{dgg} in the AGT-like  \cite{agt} framework however the explicit operator corresponding to this composite state at the Liouville side has not been constructed.

In the second part of the paper  we use the physical realization of the torus knots as the 't-Hooft loops in the $\Omega$-deformed theory. The  domain wall in this composite solution has the $S^3_b$ worldvolume hence we get the conventional geometrical framework for knot invariants.
The trajectory of the monopole at $\frac{\epsilon_1}{\epsilon_2}=\frac{p}{q}$ in the Euclidean space is identified with the torus knot.  This picture fits with the approach when
the knot invariants and homologies  can be derived from the counting of the  solutions
to the particular BPS equations  in $4d$ supersymmetric gauge theory on the interval \cite{knots1, knots2} which  is the generalization of the realization of the knot invariants  in terms of the Wilson loops in CS theory.
The knot in terms of $4d$ gauge theory is localized at the boundary of the interval  providing  the particular boundary conditions and is introduced by hands. Contrary in this paper the form of the knot is selected dynamically.  Remark that some other recent interesting results concerning the torus knots can be found in \cite{torus1,torus2,torus3}.

The relation between the torus knot invariants and the particular integrable systems of Calogero type 
at rational coupling constant allows to interpret
the knot data in terms of the auxiliary quiver gauge $2d$ theory or $3d$ theory with nontrivial boundary conditions  in the internal space.
This auxiliary gauge theory has nothing to do with original $\Omega$-deformed abelian gauge theory in $\mathbb R^4$. In fact we have in mind the AGT-like picture which has been elaborated for the hyperbolic knots in  3d/3d case  \cite{gukov2,yamazaki}.
In that case the geometry of the knot complements provides the information about the matter content and superpotentials in the dual ``physical'' SUSY 3d quiver gauge theory in the ``coordinate space''.  However
in the torus knot  case considered in this paper  the logic is a bit different. First, the torus knots are not hyperbolic  and  secondly  the knot now is located in the ``physical'' space hence its invariants  encode the information about  the internal  ``momentum space''. Usually the interpretation is opposite although the meaning of ``coordinate'' and ``momentum'' spaces in this context is a bit arbitrary.

It turns out that the relation between the torus knot invariants and the quantum integrable Calogero systems \cite{G11,G12} is useful to get the AGT-like dual representation for the knot invariants. To this aim the brane picture behind the trigonometric Ruijsenaars-Schneider (RS) model developed in \cite{gk} can be used. The rational  quantum Calogero model can be considered as the particular degeneration of RS model and is described via the $2d$ quiver gauge theory or $3d$ gauge theory on the interval in the internal space. It is crucial that the number of the particles coincides with parameter of the knot $q$. The classical rational Calogero system is dual to the quantum Gaudin model \cite{Givental,MTV1,GZZ} hence we could look how the interpretation of the knot invariants in terms of the quantum Calogero model gets translated into the quantum Gaudin side.
To this aim we have to consider the generalization of the  quantum-classical (QC) correspondence between the pair of the integrable systems to the quantum - quantum (QQ) correspondence.

The meaning of rational coupling constant in Calogero model needs some care. The point is that when both dual systems are quantum we have two different Planck constants for the Calogero and Gaudin models, say $\hbar_{Cal}$ and $\hbar_{Gaud}$. It was shown in \cite{GZZ} that $\hbar_{Gaud}$ equals to the coupling constant in the classical Calogero model. Since in the quantum Calogero Hamiltonian the classical coupling enters in the product
with the inverse $\hbar_{Cal}$ we could say that either $\hbar_{Cal}=1$ and the coupling is rational or both
Planck constants are integers,

\begin{equation}
    \hbar_{Cal}=p, \qquad \hbar_{Gaud}=q.
    \label{hbar}
\end{equation}
One could also multiply the right--hand side of these equations by some common parameter since only their  ratio matters here.
It will be argued that under the bispectrality transformation two Planck constants
get interchanged $\hbar_{Cal}\leftrightarrow \hbar_{Gaud}$. Note that physically the Planck constant corresponds to the flux of the $B$ field in the phase space
of the Hamiltonian model. We shall argue that the integers $(p,q)$ are equal to the numbers of branes hence 
they could be considered as the sources of the effective $B$ field.

The key observation
is the identification of the Dunkl operator for the quantum Calogero model with the quantum KZ equation for the Gaudin model  \cite{veselov,etingof,matsuo,cherednik}.Using the QQ correspondence for the Calogero-Gaudin pair we can suggest the counting problem for the torus knot invariants at the quantum Gaudin side.
Let us emphasize that Gaudin system considered in this paper  in terms of  the gauge theory in the internal space
in different from the Gaudin model discussed in \cite{knots2} in the physical space in the context of the calculation of the knot invariants. The third realization of the knot invariants concerns the dual Gaudin model obtained via the bispectrality transformation when the inhomogeneities get substituted by the twists \cite{harnad, MTV3}. Now the counting problem is formulated in terms of the dual KZ equation with respect to the twists.

The paper is organized as follows. In Section 2 we reconsider the central charges in the deformed theory and clarify some issues missed in \cite{bcgk}. To complete the previous analysis, we  discuss the stringy central charges and argue that they can be unified in some sense. The stringy central charge which has been used to define the $\epsilon$-string in \cite{bcgk} and can not exist without the deformation can be unified with the central charge found long time ago in the context of $\mathcal N=1$ SYM theory \cite{gs}.  In Section 3 we consider the particular configurations involving the BPS states in the $\Omega$-background. It is shown that in the ``rational'' $\Omega$-background the monopole localized at the domain wall evolves along the torus knot. In Section 4 we make some comments concerning the similar picture in the theory with fundamental matter. In Section 5 we present some arguments along the AGT logic how the torus knot can be represented by the conformal block involving the 
degenerate operators in the Liouville theory with particular value of  the central charge. In Section 6 we discuss dualities between the Calogero system and another integrable systems and gauge theories to suggest possible frameworks for consideration of torus knot invariants. Finally, a list of open questions can be found in the last Section.

\section{Central charges in $\mathcal N=2$ gauge theory}

We discuss $\mathcal N=2$ super Yang-Mills theory in presence of $\Omega$-background in four Euclidean dimensions. The field content of the theory is the gauge field $A_m$, the complex scalar $\varphi, \bar{\varphi}$ and Weyl fermions $\Lambda^I_\alpha, \bar \Lambda^I_{\dot{\alpha}}$  in the adjoint of the $U(N)$ group. Here $m=1,\dots,4$, $I=1,2$ are $SU(2)_I$ R-symmetry index, $\alpha,\dot\alpha$ are the $SU(2)_L\times SU(2)_R$ spinor indices. To introduce $\Omega$-background one can consider a nontrivial fibration of $\mathbb R^4$ over a torus $T^2$ \cite{Nekrasov},\cite{NekrasovOkounkov}. The six-dimensional metric is:

\begin{equation}
    ds^2=2dzd\bar z+\left( dx^m+\Omega^md\bar z+\bar\Omega^mdz \right)^2,
    \label{metric_Omega}
\end{equation}
where $(z,\bar z)$ are the complex coordinates on the torus and the four-dimensional vector $\Omega^m$ is defined as:

\begin{equation}
    \Omega^m=\Omega^{mn}x_n,\qquad \Omega^{mn}=\frac{1}{2\sqrt{2}}\begin{pmatrix}0&i\epsilon_1&0&0\\-i\epsilon_1&0&0&0\\0&0&0&-i\epsilon_2\\0&0&i\epsilon_2&0\end{pmatrix}.
     \label{omega}
\end{equation}

In general if $\Omega^{mn}$ is not (anti-)self-dual the supersymmetry in the deformed theory is broken. However one can insert R-symmetry Wilson loops to restore supersymmetry \cite{NekrasovOkounkov}:

\begin{equation}
    A^I_J=-\frac12\Omega_{mn}\left( \bar\sigma^{mn} \right)^I_J d\bar z-\frac12\bar\Omega_{mn}\left( \bar\sigma^{mn} \right)^I_J dz.
    \label{Wilson}
\end{equation}

The most compact way to write down the supersymmetry transformations and the lagrangian for the $\Omega$-deformed theory is to introduce 'long' scalars \cite{nw}:

\begin{equation}
    \Phi=\varphi+i\Omega^mD_m, \qquad \bar\Phi=\bar\varphi+i\bar\Omega^mD^m,
      \label{phi_def}
    \end{equation}

    Here $D_m$ are operators of covariant derivatives. This substitution reflects the fact that the scalars $\varphi, \bar \varphi$ originate from the components of the six-dimensional gauge connection along a two-dimensional torus. The metric (\ref{metric_Omega}) makes us add the rotation operators to the gauge connection, which are inherited by the complex scalars in four dimensions.
    
Then the deformed Lagrangian reads as:

\begin{equation}
  \mathcal L_\Omega=\frac{1}{4}F_{mn}F^{mn}+D_m\Phi D^m\bar\Phi+\Lambda\sigma^mD_m\bar\Lambda-\frac{i}{\sqrt{2}}\Lambda\left[ \bar \Phi,\Lambda \right]+\frac{i}{\sqrt{2}}\bar\Lambda\left[ \Phi,\bar \Lambda \right]+\frac{1}{2}\left[ \Phi,\bar\Phi \right]^2,
  \label{lagrangian}
\end{equation}
where the R-symmetry, spinor and gauge indices are suppressed. Here and in what follows we adopt Euclidean notation of \cite{ShifmanYung} for the sigma-matrices:

\begin{equation}
    \sigma_{\alpha\dot\alpha}=(1,-i\vec \tau)_{\alpha\dot\alpha},\qquad \bar\sigma_{\dot\alpha\alpha}=(1,i\vec\tau)_{\dot \alpha\alpha}.
    \label{sigma_euc}
\end{equation}

The supersymmetry algebra reads:

\begin{eqnarray}
  \delta\Phi&=&-\sqrt{2}\zeta\Lambda,\\
    \delta\bar\Phi&=&-\sqrt{2}\bar\zeta\bar\Lambda,\\
    \delta A^m &=&-\zeta\sigma^m\bar\Lambda+\bar\zeta\bar\sigma^m\Lambda,\\
    \delta\bar\Lambda&=&\sqrt{2}\bar\sigma^m\zeta D_m\bar\Phi-i\left[ \Phi,\bar\Phi \right] \bar\zeta+\bar\sigma^{mn}\bar\zeta F_{mn},\\
    \delta\Lambda&=&-i\left[ \Phi,\bar\Phi \right]\zeta+\sigma^{nk}\zeta F_{nk}+\sqrt{2}\sigma^m\bar\zeta D_m\Phi.
   \label{susy}
\end{eqnarray}
where $\zeta^I_\alpha$, $\bar \zeta^I_{\dot \alpha}$ are the supersymmetric variation parameters.

The $\mathcal N=2$ superalgebra in four dimensions admits three types of central charges which correspond one-, two- and three-dimensional defects, namely monopoles, strings and domain walls \cite{ShifmanYung}:

\begin{eqnarray}
    \left\{ Q^I_\alpha,\bar Q_{J\dot\alpha} \right\}&=& 2P_{\alpha\dot\alpha}\delta^I_J+2(Z_{\text{string}})_{\alpha\dot\alpha}\delta^I_J,\\
    \left\{ Q^I_\alpha,Q^J_\beta \right\}&=& \varepsilon_{\alpha\beta}\varepsilon^{IJ}Z_{\text{mon}}+(Z_{\text{domain wall}})^{IJ}_{\alpha\beta}.
    \label{Z_def}
\end{eqnarray}
Our current purpose is to identify these three central charges and write the BPS equations for them. The first step is to compute the N\"other current for the supersymmetry transformation. Taking the supersymmetric variation of the Lagrangian we find:

\begin{multline}
    \delta \mathcal L=\partial_m\kappa^m,\\
  \kappa^m=\bar\zeta\bar\sigma_n\Lambda F^{mn}-\sqrt{2}\zeta\Lambda D^m\bar\Phi+\zeta\sigma_n\bar\Lambda \tilde F^{mn}-\zeta\sigma^m\bar\Lambda i\left[ \Phi,\bar\Phi \right]-2\sqrt{2}\bar\zeta\bar\sigma^{nm}\bar\Lambda D_n\Phi.
  \label{kappa}
\end{multline}
Then the anti-holomorphic part of the supercurrent reads as:

\begin{equation}
  \bar J^m=\bar\sigma^n\Lambda\left( F^{mn}+\tilde F^{mn} \right)+\bar\sigma^m\Lambda i\left[ \Phi,\bar \Phi \right]-\sqrt{2}\bar\Lambda D^m\Phi+2\sqrt{2}\bar\sigma^{nm}\bar\Lambda D_n\Phi,
  \label{currents}
\end{equation}
or, after restoration of the spinor indices \cite{ShifmanYung},

\begin{equation}
    \bar J_{\dot \alpha \beta \dot \beta}=\left( -\delta_{\dot\alpha\dot\beta}i\left[ \Phi,\bar \Phi \right]+\bar\sigma^{nk}_{\dot\alpha\dot\beta}F_{nk} \right)\Lambda_\beta+\sqrt{2}\bar\sigma^n_{\beta\dot\beta}D^n\Phi\bar\Lambda_{\dot\alpha}.
  \label{current_ind}
\end{equation}

To find the bosonic part of the central charges we take the supersymmetric variation of (\ref{current_ind}) with parameters $\bar\zeta_{\dot\alpha}, \zeta_\beta$. The supervariation responsible for monopoles and domain walls reads as:

\begin{equation}
    \frac{\bar\delta_{\dot\alpha}\bar J_{\dot{\alpha}\beta\dot{\beta}}}{\sqrt{2}}=\left( -\eta^{nk}i[\Phi,\bar\Phi]+\frac12\left(F^{nk}+\tilde{F}^{nk}\right) \right)\sigma^n_{\beta\dot{\beta}} D^k\Phi.
    \label{bar_dJ}
\end{equation}
while the supervariation responsible for strings is:

\begin{equation}
    \frac{\delta_{\beta}\bar J_{\dot{\alpha}\beta\dot{\beta}}}{\sqrt{2}}=\bar \sigma^{mn}_{\dot{\alpha}\dot{\beta}}F_{mn}i[\Phi,\bar\Phi]+\sigma^{mn}\varepsilon_{mnkl}D^k\Phi D^l\bar{\Phi}+\delta_{\dot{\alpha}\dot{\beta}}\mathcal L.
  \label{Z_string}
\end{equation}
The existence of the second term in (\ref{Z_string}) was mentioned in \cite{gs} for the $\mathcal N=1$ theory. To derive the Bogomolny equations for the solitons we add the central charges to the Lagrangian to build a complete square.

In most cases to find the tension of the defect we integrate the time component of the supervariation because the defect is assumed to be static and hence stretched along time direction. But as we will see in the following the non-trivial $\Omega$-background acts effectively as an external field which affects the motion of the defect. Hence the static configuration is not realized. We assume the worldvolume of the defect to be curved and introduce unit vectors $t^n$ tangential to the worldvolume and $n^n$ normal to the worldvolume. The fields which solve the BPS equation are considered to be independent of the directions along the worldvolume,

\begin{equation}
    t^nD^n\varphi=0.
    \label{tn_def}
\end{equation}
Then the tension of the defect is the component of the supervariation along the worldvolume integrated over the directions normal to the worldvolume. Namely, the tension of the string reads as:

\begin{equation}
    T_s=\int \delta_\beta\bar J_{\dot{\alpha}\beta\dot{\beta}}\bar\sigma^{ij}_{\dot{\alpha}\dot{\beta}}dx^idx^j=\frac{i}{\sqrt{2}}\int \left(-\frac{1}{2}\left[ \Phi,\bar{\Phi} \right]\left( F^{mn}+\tilde{F}^{mn} \right)+ \varepsilon^{mnkl}D_k\Phi D_l\bar{\Phi}\right)dx^mdx^n,
    \label{T_string}
\end{equation}
and the BPS equation which describes string is the following system:

\begin{equation}
    \left\{
    \begin{array}{l}
        (F^{mn}+\tilde{F}^{mn})n_1^mn_2^n=i\left[ \Phi,\bar\Phi \right],\\
        D_w\Phi=0,\\
        D_z\Phi=0.
    \end{array}
    \right.
    \label{BPS_string}
\end{equation}

Here $z,w$ are the complex coordinates on $\mathbb C^2 \simeq \mathbb R^4$, $z=x_1+ix_2, w=x_3+ix_4$. The second and the third equations follow from the second term in the integrand of (\ref{T_string}). This object is invariant under half supersymmetries (\ref{susy}). Note that the vectors $n_1, n_2$ in the first equation can be substituted by $t_1,t_2$ since the combination $(F^{mn}+\tilde{F}^{mn})$ is self-dual.

The tension (or mass) of the monopole can be obtained in the same fashion:

\begin{multline}
    T_m=\int \bar{\delta}_{\dot{\alpha}}\bar{J}_{\dot{\alpha}\beta\dot{\beta}}\left( \bar{\sigma}^i\sigma^j\bar{\sigma}^k \right)_{\dot{\beta}\beta}dx^idx^jdx^k=\\
    2\sqrt{2}\int\left( \varepsilon_{ijkl}i\left[ \Phi,\bar{\Phi} \right]D^l\Phi+\left( F^{ij}+\tilde{F}^{ij} \right) D^k\Phi\right) dx^idx^jdx^k,
     \label{T_mon}
\end{multline}
and the BPS equation is much alike the domain wall case:

\begin{equation}
    D^m\Phi=-\frac{1}{\sqrt{2}}i[\Phi,\bar{\Phi}]t^m+\frac1{2\sqrt{2}}\left( F^{mn}+\tilde{F}^{mn} \right)t^n.
    \label{BPS_mon}
\end{equation}

The term $[\Phi,\bar{\Phi}]$ is not seen for the static monopole hence the equation (\ref{BPS_mon}) implies the usual Bogomolny equation for the monopole.
The tension of the domain wall can also be read from the supervariation (\ref{bar_dJ}):

\begin{equation}
    T_w=\int \bar{\delta}_{\dot{\alpha}}\bar{J}_{\dot{\alpha}\beta\dot{\beta}}\bar{\sigma}^i_{\dot{\beta}\beta}dx^i=2\sqrt{2}\int\left( -\eta^{mn}i\left[ \Phi,\bar{\Phi} \right]+\frac 12\left( F^{mn}+\tilde{F}^{mn} \right) \right)D^m\Phi dx^n.
    \label{T_wall}
\end{equation}

The domain wall is the defect of codimension one, hence the scalar field which builds the wall depends only on one coordinate, $\Phi=\Phi(y)$. That means that the second term in (\ref{T_wall}) drops out because $F^{mn}$ is skew symmetric and the tension has the following form:

\begin{equation}
  T_w=-2i\sqrt{2}\int\left[ \Phi,\bar{\Phi} \right]D_y\Phi dy.
  \label{T_w}
\end{equation}
The BPS equation reads as:

\begin{equation}
    D_y\Phi=-\frac{1}{\sqrt{2}}i[\Phi,\bar{\Phi}].
    \label{BPS_wall}
\end{equation}

Before proceeding further let us discuss different types of the half-BPS boundary conditions in the four-dimensional theory.

\subsection{Dirichlet and Neumann boundary conditions}
\label{section:domain_walls}
Let us consider the theory on a product of a three-dimensional space and a half-line. We interpret the boundary of this space as a three-dimensional defect or a domain wall. Let $x_0, x_1, x_2$ be the coordinates along the three-dimensional defect and $x_3$ the coordinate along the half-line. If we impose the invariance under half of the superalgebra we are left with two possibilities \cite{gw1, gw2}:

{\it Dirichlet boundary conditions.} The Dirichlet boundary conditions imply the vanishing of the components of $F^{\mu\nu}$, $\mu, \nu=0,1,2$ parallel to the domain wall, $F|_\partial=0$. We can realize a theory with these boundary conditions as a $D3$ brane ending on a $D5$ brane. The scalar fields satisfy the Nahm equations of the type (\ref{BPS_wall}), where $y=x_3$. The domain wall we described in the previous section provides the Dirichlet boundary conditions.

In the presence of $\Omega$-deformation the $\mathcal N=2$ theory acquires a superpotential. This means that the r.h.s. of the (\ref{BPS_wall}) contains an additional term $\frac{\partial W}{\partial \Phi}$. The vev of the scalar $a$ can jump on the domain wall and the dual variable $a^D$ remains constant. The domain wall supports monopoles with mass $a^D$ on its worldsheet. The condition $F|_\partial=0$ implies that the worldline of the monopole is a circle in a plane normal to the field $F_{\mu 3}$.

{\it Neumann boundary conditions.} The Neumann boundary conditions are $S$-dual to the Dirichlet ones and imply the vanishing of the other six components of $F^{\mu\nu}$, namely $\star F|_\partial=0$. They correspond to a theory on a $D3$ brane ending on an $NS5$ brane. The scalar vev $a$ remains constant across the domain wall, but the mass of the monopole $a^D$ can jump. The charged particles on the domain wall move along circles normal to the $\star F_{\mu 3}$ vector.

Of course there can be mixed boundary conditions in presence of the external gauge field whose components satisfy the following relation:

\begin{equation}
    \star F|_\partial+\gamma F|_\partial=0.
    \label{mixed_bc}
\end{equation}

If $\gamma$ is rational, $\gamma=n/m$, the domain wall providing this boundary condition supports $(m,n)$ dyons in its worldvolume.

\section{Supersymmetric solitons in $\Omega$-background}
\label{sec:solitons}
\subsection{Monopoles}



Now our goal is to find the solutions to the BPS equations for the defects of different type. The $\Omega$-background in a sense acts as external gauge field. The one-loop part of the Nekrasov partition function is derived from a Schwinger-like computation for the particle creation in the external field {\cite{NekrasovOkounkov}}. We argue that the monopoles in the $\Omega$-background move in the same fashion as in external magnetic and electric fields.

Consider the BPS equation for the monopole (\ref{BPS_mon}). Substituting the definition of the 'long' scalar (\ref{phi_def}) we have:

\begin{equation}
    D^m\varphi+i\Omega^nF^{mn}=-\frac{i}{\sqrt{2}}\left[ \varphi,\bar\varphi \right]t^m +\frac{i}{\sqrt{2}}\left( \Omega^n D^n \bar \varphi-\bar \Omega^n D^n \varphi \right)t^m+\frac{1}{2\sqrt{2}}\left( F^{mn}+\tilde{F}^{mn} \right)t^n,
        \label{BPS_mon2}
\end{equation}


Suppose that $\left[ \varphi,\bar\varphi \right]=0$. Assuming $\epsilon_{1,2}$ real and hence $\Omega^n$ imaginary, and multiplying (\ref{BPS_mon}) by $t^m$ we get

\begin{equation}
    \sqrt{2}F^{mn}\Omega^nt^m=\Omega^nD^n\left(\bar\varphi +\varphi\right),
    \label{BPS_mon3}
\end{equation}
and multiplying instead by $\Omega^n$ we get:

\begin{equation}
    \frac{1}{\sqrt{2}}(F^{mn}+\tilde{F}^{mn})\Omega^mt^n= \left( 1-i\sqrt{2}\Omega^mt^m\right)\Omega^nD^n\left( \bar\varphi+\varphi \right).
    \label{BPS_mon4}
\end{equation}

From (\ref{tn_def}, \ref{BPS_mon3}, \ref{BPS_mon4}) it is clear that tangential vector $t^n$ to the worldline of the $\epsilon$-deformed monopole is:

\begin{equation}
    t^n=\frac{i2\sqrt{2}}R\Omega^n, \qquad R^2=8\Omega^n\bar\Omega^n=\epsilon_1^2(x_1^2+x_2^2)+\epsilon_2^2(x_3^2+x_4^2)=\epsilon_1^2r_1^2+\epsilon_2^2r_2^2.
    \label{t_mon}
\end{equation}

When the ratio $\epsilon_1/\epsilon_2$ is a rational number this worldline is a torus knot $T_{p,q}$ embedded in a squashed 3-sphere $S_b^3$. The winding number  and the squashing parameter are defined as:

\begin{equation}
    \frac pq=\frac{\epsilon_1}{\epsilon_2},\qquad b^2=\frac{\epsilon_1}{\epsilon_2}.
    \label{ratio}
\end{equation}
This configuration is quite familiar to us, namely it is the worldline of the charged particle in the presence of both electric and magnetic fields.

Let us remind the calculation of the trajectory of a charge in external gauge field to see that this is indeed the case. Suppose a charged particle of spin $\sigma$ moves in presence of parallel electric and magentic fields along say $x_1$ axis. In the Euclidean signature the Dirac operator splits into two parts and the particle moves simultaneously in two circles lying in $(x_0, x_1)$ and $(x_2,x_3)$ planes. This means that the worldline of the particle is a $T_{p,q}$ torus knot if it makes $p$ rotations in one plane and $q$ rotations in the other. The action relevant for the process is the following:

\begin{multline}
    S=\int A_\mu dx^\mu+\int mds+\pi i\sigma\left( p+q \right)=\\
    pE\pi r_1^2+qB\pi r_2^2-2\pi m\sqrt{p^2r_1^2+q^2r_2^2}+2\pi i\sigma\left( p+q \right).
    \label{action_Schwinger}
\end{multline}

Extremizing (\ref{action_Schwinger}) w.r.t. radii of the circles we obtain:

\begin{equation}
    \frac{p}{q}=\frac{E}{B}, \qquad E^2r_1^2+B^2r_2^2=m^2,
    \label{wl_Schwinger}
\end{equation}
i.e. the ratio of winding numbers of the torus knot is defined from the external field, and the knot itself is embedded in a squashed three-sphere with the squashing parameter defined also by the ratio of the external fields like in (\ref{ratio}).

\subsection{Strings}

The BPS system (\ref{BPS_string}) after the substitution of the 'long' scalar (\ref{phi_def}) is:

\begin{equation}
    \left\{
    \begin{array}{l}
        \frac12\left( F^{mn}+\tilde{F}^{mn} \right)t_1^mt_2^n=i\left( \Omega^nD^n\bar \varphi -\bar\Omega^nD^n\varphi\right),\\
        wD_w(\bar\epsilon_1\varphi)=-\frac12|\epsilon_1|^2r_1^2F_{21},\\
        zD_z(\bar\epsilon_2\varphi)=-\frac12|\epsilon_2|^2r_2^2F_{34},
    \end{array}
    \right.
    \label{BPS_string2}
\end{equation}
where

\begin{equation}
    z=x_1+ix_2, \qquad w=x_3+ix_4.
    \label{complex}
\end{equation}

It is convenient here to switch to the complex coordinates:

\begin{equation}
  \Omega^nD^n=-\epsilon_1\left( zD_z-\bar zD_{\bar z} \right)+\epsilon_2\left( wD_w-\bar w D_{\bar w} \right).
  \label{Omega_C}
\end{equation}

and in the absence of external gauge fields the system (\ref{BPS_string2}) is:

\begin{equation}
  \left\{
    \begin{array}{l}
      \left( \epsilon_1\bar z D_{\bar z}-\epsilon_2 \bar w D_{\bar w} \right)\bar\varphi- \left( \bar\epsilon_1 z D_z-\bar \epsilon_2 w D_w\right)\varphi=0,\\
        wD_w\varphi= 0,\\
        zD_z\varphi= 0.
      \end{array}
  \right.
  \label{BPS_string_3}
\end{equation}

The system (\ref{BPS_string_3}) has a very natural solution, namely:

\begin{equation}
  \varphi=\varphi(z^{\bar\epsilon_2}w^{-\bar\epsilon_1}).
  \label{phi_seifert}
\end{equation}
The surface in $\mathbb C^2$ given by the equation:

\begin{equation}
  z^pw^q=\const,
  \label{def_Seifert}
\end{equation}
is called the Seifert surface for a $(p,q)$ torus knot. The Seifert surface of a knot  is by definition the surface which has a given knot as its boundary. In the realization by embedding in $\mathbb C^2$ (\ref{def_Seifert}) we need to intersect this two-dimensional surface with a three-dimensional sphere to get the torus knot. We know that in SQED case there are abelian strings which typically end on monopoles lying on the domain walls \cite{ShifmanYung}. If the trajectory of a monopole becomes a torus knot then it is natural for the corresponding abelian string to be a Seifert surface.

Of course we do not state that the Seifert surface is the only possibility for the string worldsheet. The condition of invariance under the transformations generated by $\Omega$-background also admits strings parallel to the $z$ and $w$ planes, like the ones considered in \cite{bcgk}. But these strings cannot end on a monopole solution (\ref{t_mon}) for obvious geometric reason.

In the next subsection we find that the spherical shape of the domain wall is indeed consistent with the $\Omega$-background. Hence we can argue that the composite defects containing strings intersecting domain walls along monopoles are present in the deformed theory as well as in the undeformed one.

\subsection{Domain walls}
Now let us consider the BPS equation for the domain wall (\ref{BPS_wall}), generally speaking, in presence of the superpotential:

\begin{equation}
    \sqrt{\epsilon_1 \epsilon_2}D_y\varphi+i\Omega^m F^{my}=\sqrt{2}\Omega^nD^n\varphi+\frac{\partial W}{\partial \varphi},\qquad \varphi=\varphi(y).
  \label{BPS_wall_1}
\end{equation}
Here $\varphi$ is real, $\varphi=\bar\varphi$. This condition is quite weak: it implies that the $\Omega^n$ vector is parallel to the domain wall worldvolume. This means that the monopole lies on the domain wall. The natural suggestion for the shape of the domain wall worldvolume is a squashed three-sphere with $y$ parameter equal to the squashed radius:

\begin{equation}
  y=\sqrt{\epsilon_1^2r_1^2+\epsilon_2^2r_2^2}.
  \label{y_sol}
\end{equation}
We can realize torus knot as an intersection of (\ref{def_Seifert}) with the hypersurface $y=\const$. This means that the string intersects the domain wall along the monopole worldline. Indeed, substituting (\ref{y_sol}) into (\ref{BPS_wall_1}) we get:

\begin{equation}
    D_y \varphi +(\epsilon_1 E \epsilon_1^2 r_1^2+\epsilon_2 B \epsilon_2^2 r_2^2)=\frac{\partial W}{\partial\varphi}.
    \label{BPS_wall_2}
\end{equation}

The $E$ and $B$ fields are external gauge fields which are $F_{12}$ and $F_{34}$ components of the strength tensor. For the equation (\ref{BPS_wall_2}) to depend only on $y$ we should impose the condition on the gauge fields:

\begin{equation}
    \epsilon_1 E=\epsilon_2 B,
    \label{graviphoton}
\end{equation}

which is exactly the condition that the monopole worldline is affected by the gauge fields and by the $\Omega$--background in the same way. In the absence of the external fields, the supersymmetric configuration is described by the usual equation for the domain wall $D_y \varphi=\partial W/\partial \varphi$.

The pure $\mathcal N=2$ theory does not contain dynamical supersymmetric solitons apart from monopoles, but the theory with fundamental matter does. Let us consider SQED in presence of $\Omega$-deformation and see that the worldvolumes of the defects of different dimensions change shape in the same fashion as in the pure case.

\section{Theory with fundamental matter}

Now we add the fundamental matter to the theory. In absence of the $\Omega$-background the theory supports monopoles, abelian strings and domain walls. Let us see that the presence of supersymmetric solitons is consistent with $\Omega$-deformation if the worldvolumes of the defects are curved.

For the sake of simplicity let us consider $N_f=2$. The bosonic part of the Lagrangian reads:

\begin{multline}
    \mathcal L_\Omega^m=\frac{1}{4g^2}F_{mn}^2+\frac{1}{g^2}\left|D_m\Phi\right|^2+\left|D_mq\right|^2+\left|D^m\tilde{q}\right|^2+\frac12\left|\left( \Phi+\sqrt{2}m_i \right)q_i\right|^2+\\
    \frac12 \left|\left( \Phi+\sqrt{2}\tilde{m}_i \right)\tilde{q}_i\right|^2+    \frac{g^2}{2}\left|\tilde{q}_iq_i-N\xi \right|^2+\frac{g^2}{8}\left( |q|^2-|\tilde{q}|^2 \right)^2+2g^2|[q,\tilde{q}]|^2+\\
    \frac{1}{2g^2}\left|[\Phi,\bar{\Phi}]+g^2\left[ \bar{q},q \right]+g^2\left[ \bar{\tilde{q}},\tilde{q} \right]\right|^2.
    \label{L_m}
\end{multline}
Here $\xi$ is the coefficient in front of the Fayet-Iliopoulos D-term, $m_i$ is the mass parameter, and $i=1,2$. The masses are assumed to satisfy:

\begin{equation}
  \Delta m=m_1-m_2\gg g\sqrt{\xi}.
  \label{mass_bound}
\end{equation}

First of all the BPS equations for the monopole are not changed by the presence of the matter, hence the discussion in the previous chapter remains relevant. The issue of the strings in the $\Omega$-deformed theory with the fundamental matter was discussed in \cite{bcgk}. The string BPS equations read as:

\begin{equation}
  \left\{
    \begin{array}{l}
      \left( F^{mn}+\tilde F^{mn} \right)t_1^mt_2^n+g\left( |q|^2-\xi \right)= \left[ \Phi,\bar\Phi \right],\\
  D_zq=D_wq=0,\\
  D_z\Phi=D_w\Phi=0.
  \label{BPS_string_m}
\end{array}
\right.
\end{equation}
We see that the SQED case also admits strings whose worldsheet is the Seifert surface,

\begin{equation}
  z^{\bar\epsilon_2}w^{-\bar\epsilon_1}=\const.
  \label{Seifert_def}
\end{equation}

 Consider the domain walls and first remind the construction in the undeformed theory. It has two vacua \cite{ShifmanYung, ShifmanYung1, ShifmanYung2}, the first one is:

\begin{equation}
  \varphi=-\sqrt{2}m_1, \qquad q_1=\sqrt{\xi}, \qquad q_2=0,
  \label{vac1}
\end{equation}
and the second one is:

\begin{equation}
  \varphi=-\sqrt{2}m_2, \qquad q_1=0, \qquad q_2=\sqrt{\xi}.
  \label{vac2}
\end{equation}

The theory admits the three-dimensional defect which separates these two vacua. The tension of the domain wall is:

\begin{equation}
  T_w=\xi \Delta m.
  \label{T_w_m}
\end{equation}
The transition domain between the two vacua (\ref{vac1}, \ref{vac2}) can be described as follows. The scalar field $\varphi$ interpolates between the vacuum values in the 'thick' part of the wall of the range $R\sim \Delta m$,

\begin{equation}
  \varphi=-\sqrt{2}\left( m-\Delta m\frac{z-z_0}{R} \right), \qquad |z-z_0|<R.
  \label{phi_thick}
\end{equation}
In the narrow areas of width $O\left(\left( g\sqrt{\xi} \right)^{-1}\right)\ll R$ near the edges of the wall $z-z_0=\pm R/2$ the dependence of the scalar field $\varphi$ on $z$ ceases to be linear and comes to a plateau. The quark fields inside the narrow areas interpolate between the vacua. In the thick region inside the wall the quark field is almost given by its vacuum value and depends on $z$ exponentially. Say the $q^1$ field interpolates on the left edge of the domain wall,

\begin{equation}
  q^1=\sqrt{\xi}\exp\left( -\frac{g^2\xi}{8}\left( z-z_0-\frac{R}{2} \right)^2 \right).
  \label{q_left}
\end{equation}

Then the second quark field interpolates on the right edge of the wall and generally speaking the phase of the exponential may be different from (\ref{q_left}),

\begin{equation}
  q^2=\sqrt{\xi}\exp\left( -\frac{g^2\xi}{8}\left( z-z_0+\frac{R}{2} \right)^2+i\sigma \right).
  \label{q_right}
\end{equation}

If we switch on the $\Omega$-deformation then the BPS equation for the domain wall becomes:

\begin{eqnarray}
    \sqrt{\epsilon_1\epsilon_2}D_yq&=& \frac{1}{\sqrt{2}}\left( \Phi+\sqrt{2}m \right)q,\\
    \sqrt{\epsilon_1\epsilon_2}D_y\Phi&=& \frac{g}{2\sqrt{2}}\left( |q|^2-\xi \right).
    \label{BPS_m}
\end{eqnarray}
If we assume that the domain wall is spherical like in the pure case and $y=\sqrt{\epsilon_1^2r_1^2+\epsilon_2^2r_2^2}$ then 'long' scalar acts only by multiplication and the domain wall solution is similar to the undeformed case. The tension of the wall and the qualitative structure of the fields interpolating between the vacuum values is unaffected by the non-trivial $\Omega$-background. The only difference is the spherical geometry of the wall and the fact that now the wall can in principle interact with the gauge field.

\section{AGT conjecture for surface operators wrapping the Seifert surface}
\label{sec:AGT}

In the previous Sections the $\mathcal N=2$ theory with matter was observed to admit defects of dimensions 1, 2 and 3, the geometry of which is in one or another way connected with torus knots with winding numbers defined by the ratio of the equivariant parameters, $p/q=\epsilon_1/\epsilon_2$. The AGT conjecture suggests that there is a set of corresponding operators in the Liouville theory. Although we do not provide the reader with this set of operators, we attempt to construct an operator corresponding to the two-dimensional Seifert surface and discuss, how the polynomial knot invariants can be extracted from the AGT--dual rational Liouville theory.

Let us remind the basic ingredients of the AGT correspondence \cite{agt, agt2}. The $\Omega$-deformed four-dimensional $\mathcal N=2$ gauge theory with gauge group $\prod_{i=1}^{N_f-3} SU(2)$ with $N_f$ hypermultiplets of masses $m_i$ appears to be dual to a Liouville theory on a sphere with $N_f$ punctures in the sense that the correlator of $N_f$ primary fields with  Liouville momenta $m_i$ is equal to an integral of the full partition function squared,

\begin{multline}
    \langle V_{\alpha_0}(\infty) V_{m_0}(1) V_{\alpha_{n+1}(0)} \prod_{i=1}^{N_f-3}V_{m_i}(q_1\ldots q_i)\rangle=\\
    cf(\alpha_0)f\left( \alpha_{n+1} \right)
    \prod_{i=1}^{N_f-3}f(m_i)\int \prod_{i=1}^{N_f-3}(a_i^2da_i)\left|Z_{\alpha_0\alpha_1\dots\alpha_{n+1}}^{m_0m_1\dots m_n}(q_i)\right|^2,
  \label{agt}
\end{multline}
where $\alpha_i=Q/2+a_i, i=1,\ldots,n$, and $\alpha_0, \alpha_{n+1}, m_i$ are linear combinations of the background charge $Q$ and masses of the hypermultiplets. 

The central charge of the Liouville theory is defined by the deformation parameters,

\begin{equation}
  c=1-6Q^2, \qquad Q=b+1/b, \qquad b^2=\epsilon_1/\epsilon_2.
  \label{agt_b}
\end{equation}

The insertion of the surface operator in the four-dimensional gauge theory results in insertion of the degenerate field in the Liouville correlator \cite{agt2}. Namely, if $\epsilon_{1,2}$ parameters correspond to rotations in $z_{1,2}$ planes (where $z_1=x_1+ix_2, z_2=x_3+ix_4$), then the surface operator along the $z_1$ plane corresponds to insertion of $V_{1,2}(z)$ field and the surface operator along the $z_2$ plane corresponds to $V_{2,1}(z)$ field. How to construct the operator corresponding to Seifert surface? Although we do not know the full answer to this question, we could try to suggest a construction using the theory of knot polynomials.

The HOMFLY polynomial for a given $(p,q)$ torus knot can be calculated using the Calogero integrable system \cite{G11}. Consider a system of $q$ Calogero particles with coupling constant equal to $\nu=p/q$,

\begin{equation}
    H=\frac 12\sum_{i=1}^q \frac{\partial^2}{\partial x_i^2}+\sum_{i\neq j}\frac{\nu(\nu-1)}{\left( x_i-x_j \right)^2}.
    \label{H_Cal}
\end{equation}

The Calogero Hamiltonian (\ref{H_Cal}) can be written as a square of the Dunkl operator,

\begin{equation}
    \mathcal D_i=\partial_i+\nu\sum_{i\neq j}\frac{s_{ij}-1}{x_i-x_j}.
    \label{def_Dunkl}
\end{equation}
where $s_{ij}$ is the permutation operator. The model possesses an $\mathfrak{sl}_2$ symmetry, generated by operators $(H,K,D)$:

\begin{equation}
    H=\sum_i \mathcal D_i^2, \qquad K=\frac12\sum_i (\mathcal D_i x_i +x_i \mathcal D_i), \qquad D=\sum_i x_i^2,
    \label{sl2}
\end{equation}
where $K$ is the dilation generator, $D$ is the conformal boost. This $\mathfrak{sl}_2$ is a subalgebra of the rational Cherednik algebra \cite{cherednik}.

The HOMFLY polynomial can be computed from the action of the Cherednik algebra on the factor of the polynomial ring over the kernel of the Dunkl operator (cf. \ref{app_homfly}),

\begin{equation}
    \mathcal D_i \psi=0,\qquad i=1, \ldots, q.
    \label{Dunkl_eq}
\end{equation}

The solutions to the equation (\ref{Dunkl_eq}) and consequently to the equation $H\psi=0$ are polynomials in $x_i$. But the Calogero system admits also eigenfunctions which are rational functions of $x_i$. Hence we can write the Calogero Hamiltonian as a square of another operator,

\begin{equation}
    \tilde {\mathcal D_i}=\partial_i+ (1-\nu)\sum_{i\neq j} \frac{s_{ij}-1}{x_i-x_j}.
    \label{Dunkl_alt}
\end{equation}

Then the equation (\ref{Dunkl_eq}) can be solved by functions having negative powers of $x_i$.

We can interpret the conditions $H\psi=0$ and $\tilde {\mathcal D_i} \psi=0$ as BPZ and KZ conditions on Liouville correlators. The BPZ equation \cite{BPZ} for a correlator of fields $\varphi_i$ with dimensions $h_i$ reads as:

\begin{equation}
    \left(-\frac{3}{2(2h+1)}\frac{\partial^2}{\partial x_i^2}+\sum_{j\neq i}^N\left( \frac{1}{x_i-x_j}\frac{\partial}{\partial x_j}+\frac{h_i}{(x_i-x_j)^2} \right)\right)\langle \varphi_1 \ldots \varphi_N \rangle=0.
    \label{BPZ}
\end{equation}

To obtain the Calogero Hamiltonian (\ref{H_Cal}) we consider the set of BPZ operators on the $q$--point correlation function of $V_{1,2}$ operators. The dimension of $V_{1,2}$ operator is 

\begin{equation}
    h=-\frac{3b^2}{4}-\frac12.
    \label{h_degenerate}
\end{equation}

The BPZ equations for this correlator look as follows:

\begin{equation}
    \left(b^{-2}\partial_i^2 +\sum_{j\neq i}\frac{\partial_j}{x_j-x_i}+\sum_{j\neq i}\frac{-\frac{3b^2}{4}-\frac12}{\left( x_i-x_j \right)^2}\right)\langle V_{1,2}(x_1)\ldots V_{1,2}\left( x_q \right)\rangle=0.
    \label{BPZ_surface}
\end{equation}

Making a substitution (which amounts to decoupling from a correlator a factor of $\prod_{i\neq j}(z_i-z_j)^{\left(\frac{b^2}{2}\right)}$):

\begin{equation}
    \partial_i \rightarrow \partial_i-\frac{b^2}{2}\sum_{j\neq i}\frac{1}{x_i-x_j},
    \label{sub_BPZ}
\end{equation}
and summing all the equations in the system (\ref{BPZ_surface}) we arrive exactly to the equation on the Calogero wavefunction with zero energy:

\begin{equation}
    \left(\sum_i^q \frac{\partial^2}{\partial x_i^2}-\sum_{i\neq j}^q\frac{b^2(b^2+1)}{(x_i-x_j)^2}\right)\langle V_{1,2}(x_1)\ldots V_{1,2}(x_q)\rangle=0.
    \label{BPZ_Cal}
\end{equation}
Instead of this product, we could consider a product of $p$ $V_{2,1}$ operators and arrive to the same answer with $b\leftrightarrow b^{-1}$. The
similar relation between the conformal blocks in the conformal theory and the Calogero wave functions with different energies has been found 
in \cite{cardy}.

The operator $\prod V_{1,2}(x_i)$ can be considered as a partial answer to the question about the Liouville counterpart of the surface operator lying along a Seifert surface. Indeed, we want to construct from operators $V_{1,2}(z_2)$ which corresponds to a plane along $z_1$ and $V_{2,1}(z_1)$ which corresponds to a plane along $z_2$ an operator corresponding to a surface:

\begin{equation}
    z_1^{q}=z_2^p, \qquad p/q=b^2.
    \label{seifert}
\end{equation}

From a brane construction of the two-dimensional defects it is natural to suggest that the Liouville counterpart of this surface operator contain $q$ copies of $V_{1,2}$ operator or equivalently $p$ copies of $V_{2,1}$ operator. Hence we can make a conjecture that the AGT correspondence maps a two-dimensional defect along the Seifert surface into a product of degenerate fields, and the description of the torus knot invariants in terms of Calogero eigenstates proposed in \cite{G11, G12} arises from a consideration of the expectation value of the corresponding Liouville operator.

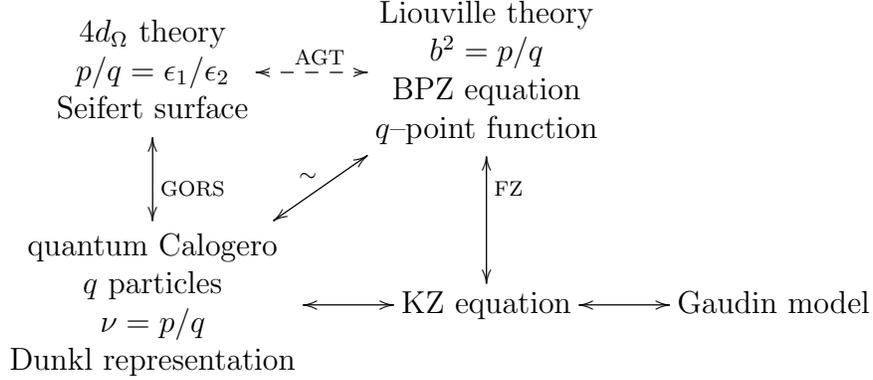
\begin{figure}
    \centering
    \begin{equation*}
        \xymatrix{
            {\begin{matrix}4d_\Omega \text{ theory}\\p/q=\epsilon_1/\epsilon_2\\\text{Seifert surface}\end{matrix}}\ar@{<-->}[r]^{\text{AGT}}\ar@{<->}[d]^{\text{GORS}}&{\begin{matrix}\text{Liouville theory}\\b^2=p/q\\\text{BPZ equation}\\{q \text{--point function}}\end{matrix}}\ar@{<->}[d]^{\text{FZ}}\ar@{<->}[ld]_\sim&\\
            {\begin{matrix}\text{quantum Calogero}\\q \text{ particles}\\\nu=p/q\\ \text{Dunkl representation}  \end{matrix}}\ar@{<->}[r]&\text{KZ equation}\ar@{<->}[r]&{\text{Gaudin model}}
        }
        \end{equation*}
        \label{fig:agt}
        \caption{AGT-like construction for torus knot invariants. Here GORS stands for the technique developed in \cite{G12} for computing the torus knots invariants in terms of the Calogero model (see also appendix \ref{app_homfly}), AGT stands for AGT mapping which presumably relates Seifert surface on the four-dimensional side with a $q$-point correlator of degenerate Liouville fields, FZ for Fateev-Zamolodchikov \cite{FZ} correspondence, $\sim$ denotes formal coincidence of the BPZ equations with quantum Calogero system. The composition of two horizontal arrows in the second line is interpreted as a manifestation of the quantum--quantum correspondence. }
\end{figure}

The equation $\tilde{\mathcal D_i}\psi=0$ can be considered as a Knizhnik-Zamolodchikov equation \cite{KZ} in a corresponding WZW model. Indeed, in \cite{FZ} it was stated that the Liouville correlators with insertion of a degenerate fields $V_{1,2}$ or $V_{2,1}$ are equal to certain correlators in $SU(2)$--WZNW model.  

The relation between KZ operator and Dunkl operator with integer coupling constant $\nu\in \mathbb Z$ was noted in \cite{veselov}. We can consider a KZ equation for an $n$-point correlator,

\begin{equation}
    \partial_i \psi=\left(\nu\sum_{j\neq i} \frac{s^{ij}}{x_i-x_j}+ \lambda^{i} \right)\psi
    \label{Gaudin_eq}
\end{equation}
where  $\Psi$ takes values in the tensor product  $V^{\otimes n}$  and $\lambda^i $ is the operator acting
as $\lambda$ on $i$-th factor and identically on the others. If $\mathrm{dim} V=n$, then we can decompose $\psi$ as a sum over permutations of indices,

\beq
\psi= \sum_{\sigma\in S_n} \Phi_{\sigma} e_{\sigma},  \qquad e_{\sigma} =  e_{\sigma(1)}\otimes \ldots \otimes   e_{\sigma(n)},
\eeq

The operator $s_{ij}$ can be written as a certain linear operator on the space of vectors $\Phi_\sigma$ and can be considered as a tensor product of $SU(2)$ generators $t_a\otimes t_b$ entering the KZ equation. Certain combinations of $\Phi_\sigma$,

\begin{equation}
    \Psi^{Cal}_\nu=\sum_{\sigma\in S_n} \Phi_{\sigma},\qquad \Psi^{Cal}_{\nu+1}=\sum_{\sigma\in S_n}\mathrm{sign}(\sigma)\Phi_\sigma,
   \label{psi_Cal}
\end{equation}
are the eigenfunctions of Calogero with coupling constants $\nu$ and $\nu+1$ respectively. There is an analogous construction for a case of general rational coupling \cite{etingof}.

\section{Quantum--classical duality in integrable systems}

\subsection{QC duality between Gaudin and  Calogero models}

\label{section:GMTV}

In the Section \ref{sec:solitons} we argued that the worldlines of monopoles in $\Omega$--background form torus knots, and that $\Omega$--background generically admits two-dimensional defects forming a Seifert surface. In the Section \ref{sec:AGT} we conjectured that the torus knot invariants can be extracted from certain correlators in the Liouville theory which are related to surface operators in four-dimensional theory by the AGT correspondence. The polynomial invariants are computed through the Calogero model arising from the BPZ set of equations on the correlation function. The key step in the computation is the expression of the original Calogero problem in terms of Dunkl operators. The eigenvalue problem for Dunkl operators formally coincides with the eigenvalue problem for the quantum Gaudin system. 

The classical Calogero model is known to be dual in a certain sense to the quantum Gaudin model \cite{GZZ}. Conjecturally, this duality can be lifted to quantum--quantum level. The elements of this quantum--quantum correspondence have been considered in \cite{matsuo,cherednik,veselov,etingof}. In this section we shall make some preliminary work concerning this issue postponing detailed discussion   for the separate study.

\begin{figure}
    \centering
    \includegraphics[height=70pt, width=100pt]{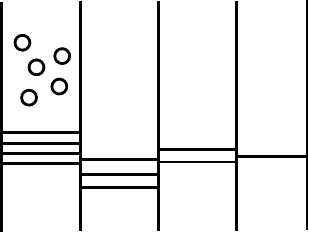}
    \caption{The Bethe ansatz equation for the Gaudin model (\ref{BAE_Gaudin}) is written for the quiver which contains a flavor group only in the first node. See also chapter \ref{section:branes}.}
    \label{fig:nesting}
\end{figure}

We follow the explicit construction of the classical Calogero--quantum Gaudin  QC duality provided by \cite{GZZ}. Consider the Calogero Lax operator,

\begin{equation}
    L_{ij}^{Cal}=\delta_{ij}{\dot x}_i+\nu \frac{1-\delta_{ij}}{x_i-x_j}, \qquad i,j=1,\ldots,N.
    \label{Lax_Cal}
\end{equation}
To get the Bethe ansatz equations for the Gaudin model, we consider the intersection of two Lagrangian submanifolds in the Calogero phase space, namely we fix the spectrum of the Lax operator and  all coordinates. If we identify the classical Calogero coupling with the Gaudin Planck constant,

\begin{equation}
    \nu=\hbar_{Gaud},
    \label{h_Gaudin}
\end{equation}
the velocities of the Calogero model are equal to the Gaudin Hamiltonians evaluated at the solutions to the Bethe equations,

\begin{equation}
    \dot{x}_j=\frac 1\hbar H_j^G\left( {\bf x}_N, {\bf \mu}^1_{N_1},\ldots, {\bf \mu}^{q-1}_{N_{q-1}} \right), \qquad j=1,\cdots,q,
    \label{vel}
\end{equation}
where integers $q$ is the number of sites and $N_i$ are the number of Bethe roots at the $i$-th level of nesting,

\begin{equation}
    N\ge N_1\ge N_2 \ge \ldots \ge N_{p-1}\ge 0,
    \label{eigen_Gaudin}
\end{equation}
The  spectrum of the Calogero Lax operator consists of $n$ different eigenvalues,

\begin{equation}
    \mathrm{Spec}L^{Cal}({\bf \dot x}_N, {\bf x}_N, \nu)=( \underbrace{v_1,\ldots,v_1}_{\mbox{\small{$N-N_1$}}}, \underbrace{v_2,\ldots,v_2}_{\mbox{$N_1-N_2$}}, \ldots, \underbrace{v_p,\ldots,v_p}_{\mbox{$N_{p-1}$}}).
    \label{spec}
\end{equation}
The nested Bethe ansatz equations for the Gaudin system are the following:

\begin{equation}
    v_b-v_{b+1}+\delta_{1b}\sum_{k=1}^q\frac{\hbar_{Gaud}}{\mu^b_\beta-x_k}=-\sum_{\gamma=1}^{N_{b-1}}\frac{\hbar_{Gaud}}{\mu^b_\beta-\mu^{b-1}_\gamma}+2\sum_{\gamma\neq\beta}^{N_b}\frac{\hbar_{Gaud}}{\mu^b_\beta-\mu^b_\gamma}-\sum_{\gamma=1}^{N_{b+1}}\frac{\hbar_{Gaud}}{\mu^b_\beta-\mu^{b+1}_\gamma}.
    \label{BAE_Gaudin}
\end{equation}
where $\mu$ variables correspond to the Bethe roots and $v$ variables to the twists.
The equation (\ref{BAE_Gaudin}) fixes all the impurities to be on the first level of nesting, see fig. \ref{fig:nesting} for the brane picture. We see that the relation (\ref{vel}) is the classical limit of the Dunkl equation (\ref{Dunkl_eq}).
The same construction is valid for the Ruijsenaars-XXX chain correspondence.

The equation (\ref{Dunkl_eq}) is equivalent to the KZ equation for the $SL(p)$ $q$-point conformal block in Liouville system and to the KZ equation involving Gaudin Hamiltonians \cite{etingof}. Here the coupling of the quantum Calogero system is identified with the $b$ parameter in the Liouville theory, $\nu=b^2$. In the paper \cite{etingof} it was explicitly demonstrated how the finite-dimensional representation of the Cherednik algebra can be
constructed in terms of the solution to KZ equation. Hence the torus knot invariants can be expressed in terms of characters of the Cherednik algebra realized on the conformal blocks in the rational models.

\subsection{QC duality via branes}
\label{section:branes}

To get link with the previous physical realization of the torus knots it is useful to consider the brane picture behind
the Calogero system and spin chain. Remarkably
the duality between them has been was identified as the correspondence between the quiver $3d$  $\mathcal N=2^*$ gauge theory and $\mathcal N=2^*$  $4d$ gauge theory on the interval \cite{gk}. The integrable data are encoded in the structure of quiver in the $3d$ theory and in  the boundary condition for the $4d$ gauge theory with $\mathcal N=2$ SUSY with
$\mathbb R^2\times S^1 \times L$ geometry.  It is assumed that there are different boundary conditions at the ends of the interval.

Consider  $M$ parallel  $NS5$ branes  extended in $(012456)$, $N_i$ $D3$ branes extended
in $(0123)$ between $i$-th and $(i+1)$-th $NS5$ branes, and $M_i$ $D5$ branes extended
in $(012789)$ between $i$-th and $(i+1)$-th $NS5$ branes (see table (\ref{table:branes})). From this brane configuration
we obtain the $\prod_{i}^{M} U(N_i)$ gauge group on the $D3$ branes worldvolume with
$M_i$ fundamentals for the $i$-th gauge group. The distances between the $i$-th and $(i+1)$-th
$NS5$ branes yield the complexified gauge coupling for $U(N_i)$ gauge group while
the coordinates of the $D5$ branes in the $(45)$ plane correspond to the masses
of fundamentals. The positions of the $D3$ branes on $(45)$ plane correspond to the coordinates on the
Coulomb branch in the quiver theory.
The additional $\Omega$ deformation reduces the theory with $\mathcal N=4$ SUSY
to the $\mathcal N=2^*$ theory. It is identified as $3d$ gauge theory when the distance between $NS5$
is assumed to be small enough. In what follows we assume that one coordinate
is compact that is the theory lives on $\mathbb R^2\times S^1$.

\begin{figure}
\centering
\begin{tabular}{|c|c|c|c|c|c|c|c|c|c|c|}
    \hline
    &0&1&2&3&4&5&6&7&8&9\\
    \hline
    D3&$\times$&$\times$&$\times$&$\times$&&&&&&\\
    \hline
    NS5&$\times$&$\times$&$\times$&&$\times$&$\times$&$\times$&&&\\
    \hline
    D5&$\times$&$\times$&$\times$&&&&&$\times$&$\times$&$\times$\\
    \hline
\end{tabular}
\caption{Brane construction of the $3d$ quiver theory.}
\label{table:branes}
\end{figure}

\begin{figure}
    \centering
    \includegraphics[height=70pt, width=250pt]{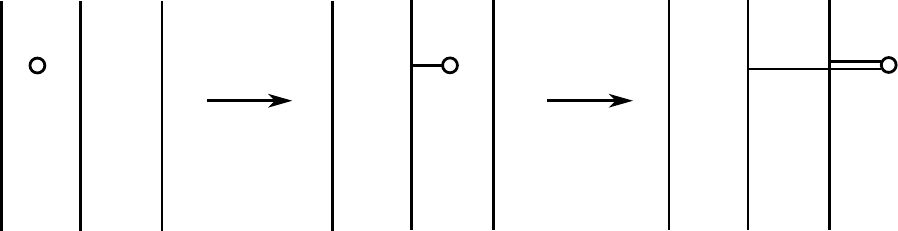}
    \caption{Hanany-Witten transformation. Here vertical lines are $NS5$ branes, horizontal lines are $D3$ branes, and circles are $D5$ branes. When a $D5$ brane is moved through a sequence of $NS5$ branes the linking number between them is conserved hence additional $D3$ branes appear.}
    \label{fig:HWmove}
\end{figure}

The other way to look on this construction is to consider four-dimensional theory on the interval between two domain walls of the Neumann/Dirichlet type as in the chapter \ref{section:domain_walls}. Performing Hanany-Witten transformations \cite{hanany_witten} (see fig. \ref{fig:HWmove}) we can place all the $D5$ branes to the left of the $NS5$ branes.
Hence now we have a $U(Q)$ four-dimensional gauge theory placed between Neumann boundary conditions provided by $M$ $NS5$ branes and Dirichlet boundary conditions provided by $N=\sum_j  M_i$ D5 branes
\begin{equation}
    Q= \sum_{j=1}^{p} jM_j.
    \label{eq:Q}
\end{equation}
The information about the $3d$ quiver is now encoded in the boundary conditions in the $4d$ theory via embedding $SU(2)\rightarrow U(Q)$ at the left and right boundaries \cite{gw1,gw2,nw}.

The mapping of the gauge theory data into the integrability framework goes as follows. In the NS limit of the $\Omega$-deformation the twisted superpotential in $3d$ gauge theory
on the $D3$ branes gets mapped into the Yang-Yang function for the $XXZ$ chain \cite{ns1,ns2}. The minimization of the superpotential yields the equations
describing the supersymmetric vacua and in the same time they are the Bethe ansatz equations for the $XXZ$ spin chain. That is $D3$ branes
are identified with the Bethe roots which are distributed according to the ranks of the
gauge groups at each of the $p$ steps of nesting $\prod_{i}^p U(N_i)$. Generically the number
of the Bethe roots at the different levels of nesting is different. The distances between
the $NS5$ branes define the twists at the different levels of nesting while the
positions of the D5 branes in the $(45)$ plane correspond to the inhomogeneities
in the $XXZ$ spin chain. To complete the dictionary recall that the anisotropy of the $XXZ$ chain
is defined by the radius of the compact dimensions while the parameter of the $\Omega$
deformation plays the role of the Planck constant in the $XXZ$ spin chain.

The interpretation of quantum-classical duality we are interested in goes as follows \cite{gk}. We interpret it as the duality between the $3d$ quiver theory and the $4d$ theory on the interval.
The moduli of the vacua in the $\mathcal N=2^*$ $4d$ $U(Q)$ gauge  theory
are known to be parameterized
by the $U(Q)$ flat connections on the torus with one marked point with particular
holonomy determined by the deformation parameter. This is exactly the
description of  phase space of the trigonometric RS model with $Q$ particles \cite{gornek}. Now the boundary conditions fix the two Lagrangian submanifolds
in this space. The Dirichlet boundary fixes the coordinates while the Neumann boundary
fixes the eigenvalues of the Lax operator. We arrive at the picture of
intersection of two Lagrangian submanifolds in the trRS model we worked with.
This picture has been developed for the first time in \cite{nrs}. For the application to the torus knot invariants we shall need the non-relativistic limit of this correspondence, namely Calogero-Gaudin correspondence corresponding to the small radius of the circle. Hence we arrive just to the picture described in the chapter \ref{section:GMTV}.

\begin{figure}
    \centering
    \includegraphics[height=70pt, width=180pt]{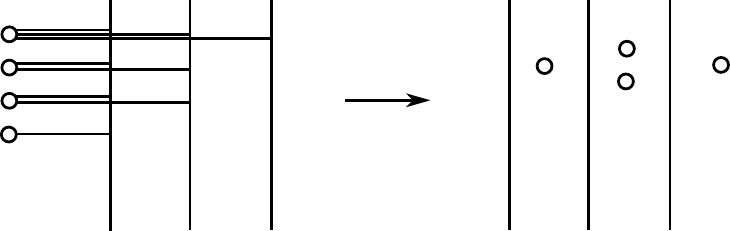}
    \caption{A sequence of Hanany-Witten moves can transform the brane configuration describing $4d$ theory on the interval (left) to the brane configuration consisting only of $NS5$ and $D5$ branes (right) if {\it admissibility} is satisfied. Here on the picture $p=3, q=4, M_1=4, M_2=1$. }
    \label{fig:GMTV_branes}
\end{figure}

The Hanany-Witten transformation allows to simplify the combinatorial problem of enumerating all the configurations consisting only of $NS5$ branes and $D3$ branes in the following way. Let us use the notation of the chapter \ref{section:GMTV}. Then we are considering the quiver defined by $p$ $NS5$ branes and $q$ $D5$ branes, where $p\le q$. The rank of the gauge group at $b$-th node of the quiver is given by $N_b$ and the rank of the flavor group by $M_b$, $b=1,\ldots,p-1$, $\sum M_b=q$. We also assume $N_0=N_p=0$. The system in chapter \ref{section:GMTV} corresponds to the case $M_2=M_3=\ldots=0$.

Following \cite{gw1} we impose on the set of numbers $(N_b,M_b)$ a certain restriction which we will call the {\it admissibility} condition which ensures a nice RG flow for the theory in the IR,

\begin{equation}
    2N_b\le N_{b-1}+N_{b+1}+M_b.
    \label{adm}
\end{equation}

Remarkably the same inequality arises in \cite{krichever} when the nested Bethe ansatz equations for an elliptic system are studied. This inequality turns into equation in the elliptic case and is a certain property of zeroes of sigma-function. When the limit to trigonometric or rational case is taken, the equation degenerates to (\ref{adm}) with $N_b=0$.

If we want to enumerate all the configurations of $D3$ branes satisfying the {\it admissibility} then we can adopt the Hanany-Witten move to simplify this combinatorial problem.
Suppose that what we consider is $q$ $D5$ branes distributed somehow between $p$ $NS5$ branes or to the left of them. There are no $D3$ branes present. Hence this configuration is always {\it admissible}.
Suppose that we make a Hanany-Witten transformation and place all the $D5$ branes to the right of the $NS5$ branes (see fig. \ref{fig:GMTV_branes}). Then we have only $D3$ branes left between the $NS5$ branes. However the configuration is still {\it admissible} since the Hanany-Witten move respects the (\ref{adm}) condition.

Now the claim is the following: every {\it admissible} configuration consisting of $D3$ branes distributed between $p$ $NS5$ branes and with $q$ $D5$ branes to the right of them can be transformed by a chain of Hanany-Witten moves to the configuration containing no $D3$ branes and $q$ $D5$ branes distributed between $p$ $NS5$ branes. This is really simple: one can easily show that in the latter configuration the number of these $D5$ branes at $b$-th node $K_b$ is given by:

\begin{equation}
    K_b=N_{b-1}+N_{b+1}-2N_b.
    \label{K_b}
\end{equation}

If we want this to be positive we impose {\it admissibility}.
Hence the problem of finding all the {\it admissible} configurations consisting of $D3$ branes is reduced to the problem of distributing $D5$ branes between and to the left of the $NS5$ branes.
From (\ref{K_b}) it follows immediately that the number of the $D5$ branes lying to the left of the $b$-th node is $N_b-N_{b+1}$. This means that the degeneration of the spectrum of the Calogero Lax operator counts the number of the $D5$ branes located to the left of each $NS5$ brane.
Of course we can draw the condition (\ref{eigen_Gaudin}) as an $p\times q$ Young diagram and then we are left with counting the Young diagrams satisfying {\it admissibility}.
Perhaps the problem of the calculating of the torus knot invariants can be reduced to sum over the brane configurations with some weight. This problem deserves separate consideration.

\subsection{Torus knots in various frameworks}
\label{section:combining}

Hence we arrive at the following picture. The torus knot in the $4d$ Euclidean space is represented by the trajectory of the
monopole in the $\Omega$-background  localized at the domain wall. The invariants of the knot are described in terms of the quantum Calogero model, which at first glance is consistent with the AGT conjecture. The classical Calogero model is connected with a quantum Gaudin model, which can be interpreted as a classical limit of the Dunkl representation for the Calogero model. Hence we propose the question about the meaning of knot invariants in various integrable models.

The Calogero model with rational coupling $\nu=p/q$ describes the vacuum manifold for the particular
gauge theory. This theory can be considered as the limit of $SU(q)$ $4d$ gauge theory on $\mathbb R^2\times S^1\times I$ at small radius of the circle and nontrivial boundary conditions imposed by the $q$ $D5$  and $p$ $NS5$ branes.

\begin{figure}
    \centering
    \begin{equation*}
            \xymatrix{
                               {\begin{matrix} \text{classical Calogero} \\ \nu=p \\ q \text{ particles} \end{matrix}}\ar@{<->}[rr]^{\text{QC}} &&{\begin{matrix} \text{Gaudin}\\ \hbar=p\\ p \text{ levels of nesting}\\ q \text{ inhomogeneities} \end{matrix}}\\
               &&&\\
                {\begin{matrix}\text{trigonometric}\\\text{Ruijsenaars}\end{matrix}}\ar[uu]^{\begin{matrix}R\to 0\\\epsilon\to 0\end{matrix}}\ar@{<->}[rr]^{\text{QC}}\ar@{<->}[d]&&{\begin{matrix}XXZ \text{ spin chain}\end{matrix}}\ar[uu]^{\begin{matrix}R\to 0\\\epsilon\to 0\end{matrix}}\ar@{<->}[d]^{\text{NS}}\\
                {\begin{matrix}4d \text{ theory on interval}\\ p\, NS5 \text{ branes}\\ q\,D5 \text{ branes}\end{matrix}}\ar@{<->}[rr]^{\text{HW move}} && {\begin{matrix}3d \text{ quiver}\\ p\, NS5 \text{ branes}\\ q\,D5 \text{ branes}\end{matrix}}
            }
     \end{equation*}
    \label{fig:dualities}
    \caption{Various dualities between quantum and classical integrable systems and gauge theories. The quantum--quantum version of this correspondence can presumably describe polynomial knot invariants.  QC stands for quantum--classical duality, NS for Nekrasov--Shatashvili correspondence, HW for Hanany--Witten transformation.}
\end{figure}
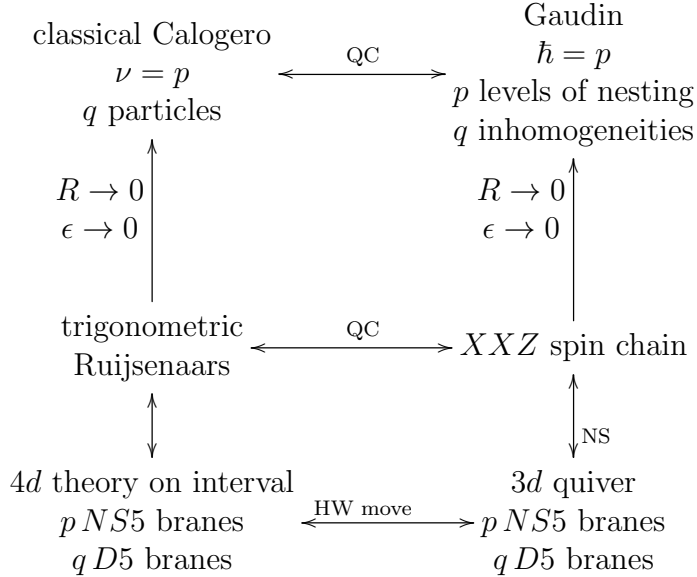

This  gauge theory can be related via the HW move to the quiver $3d$ gauge theory at small radius which can be effectively considered as the quiver $2d$ theory. The Hilbert space of this theory can be described by the twisted $SL(p) $ Gaudin spin chain
at $q$ sites. The way to extract the torus knot invariants deals now with the solutions to the KZ equation with respect to the inhomogeneities. The Planck constant in the Gaudin model is identified with the number of $NS5$ branes while the Kac-Moody level involving the KZ equation is identified with the ratio $p/q$. This fits with the similar interpretation of the parameters of $\Omega$-background in the AGT  correspondence.

The third way to consider the same problem appears upon the application of the bispectrality at the Gaudin side \cite{harnad, MTV3}. Indeed
in this case one considers the $SL(q)$ Gaudin model on $p$ sites when inhomogeneities attributed to $D5$ branes and twists attributed to $NS5$ get interchanged. The KZ equation with respect to the position of the marked points gets substituted by the dual KZ with respect to the twists.

The knot invariants have different interpretation in all these cases. In the Calogero system they count the finite-dimensional part of the spectrum with respect to two gradings. One grading corresponds to the Cartan in the $\mathfrak{sl_2}$ while the second accounts for the representation of the symmetric group. At the Gaudin side we consider the KZ equation and take into account the emergence of the finite dimensional representation of the Cherednik algebra at the rational Kac-Moody level \cite{etingof}. Recall that $\mathfrak{sl_2}$ above is just subgroup of Cherednik algebra. Then roughly speaking we consider the character of this finite-dimensional representation in terms of solution of KZ. The bispectrality can be applied to this KZ equation as well so we can consider the similar counting problem for the inverse Kac-Moody level. From the point of view of the torus knots, the bispectrality acts as the mirror reflection $p\leftrightarrow q$. If we restore the Planck constant at the Calogero site the bispectrality interchange the Planck constants of quantum Calogero and quantum Gaudin models.

Let us emphasize that the identification of the torus  knot invariants in terms of the Hilbert space of the Calogero model is relatively clear from the different viewpoints. On the other hand the dualities between the integrable systems discussed above suggest the new realization of the knot invariants in terms of the Hilbert space of the pair of the Gaudin models related by bispectrality. We have not present the precise realization of the torus knot invariants at the Gaudin side and hope to discuss this issue elsewhere. This problem  is actually closely related with the representation of the string wrapped on the Seifert surface at the Liouville side of the AGT correspondence discussed at Section \ref{sec:AGT}.

\section{Discussion}

In this paper the exact solutions describing the particular composite supersymmetric solitons in the $\Omega$-deformed $\mathcal N=2$ theory were found. In presence of the $\Omega$-background the particle-like half-BPS solitons move along along a torus knot embedded in a squashed three-sphere. For the monopoles the parameter of the squashing and the ratio of the winding numbers of the knot are connected with the ratio of the deformation parameters $\epsilon_1/\epsilon_2$. The monopole is bound to the worldsheet of the domain wall and presumably can be interpreted as the end of the solitonic string.

Given the physical realization of the torus knot we have discussed  the realization of its HOMFLY invariants in the different frameworks. In particular we shall exploit the relation between the torus knots and quantum Calogero model at rational coupling to formulate the meaning  of  the  knot invariants purely in terms of the integrable model. The dualities between the integrable models imply several interesting realizations of the knot invariants, for instance in terms of the KZ equations at rational level corresponding to the minimal models. In this case the quantum-quantum duality between Calogero and Gaudin models plays the key role. The brane setup behind the integrable models of Calogero and spin chain type helps to clarify some geometrical aspects.

It is clear that there are many issues to be  answered and we list a few below.
\begin{itemize}

\item It would be very interesting to clarify the relation of our picture with other representations of the torus knot invariants.
One approach concerns their realization as the integral of the proper observables over the abelian instanton moduli space in the sector with fixed instanton number \cite{gorneg}. The second realization concerns their interpretation as the partition function of the surface operator carrying the magnetic flux \cite{holland}. This  partition function is saturated by the
instantons trapped by the surface operator.

\item The described approach hints that the torus knot invariants can be obtained by means of enumeration of the solutions to the BAE or equivalently of enumeration of some brane configurations. It would be interesting to clarify this connection and the role of brane moves on the knot side of the correspondence.

\item Another interesting problem is to perform the summation  over instanton number within the framework of \cite{gorneg} where a $(p,q)$ torus knot superpolynomial is represented as some integral over the $q$-instanton moduli space.

\item It is natural to generalize the present analysis to the RS model and try to formulate the torus knot superpolynomial through the spectrum of the quantum RS model. The identification of the knot homologies purely in terms of the Hilbert space of  Hamiltonian system is expected as well.

\item   It is interesting to look for  the possible relation between the algebraic sector in the quasiexactly solvable models
and torus knots. Presumably it can be  interpreted in terms of the spectral curves of the corresponding Hamiltonian systems.

\item There are some additional structures which appear in the stable limit $p,q\rightarrow \infty$ of the torus knot. This limit has different interpretations in all approaches mentioned. In the initial 4D gauge theory it corresponds to the strong external fields. In the Calogero model the number of particles tends to infinity simultaneously with the coupling. This limit is usually described in terms of the collective field theory. At the Gaudin side the number of sites tends to infinity therefore one has to discuss the thermodynamical limit. Finally, in the brane picture it is the  limit where  the number of $NS5$ branes and/or  
the number of D5 brane tends to infinity. It would be interesting to match these pictures.

\item The torus knot represented by the composite BPS state is the Euclidean configuration possessing a negative mode.
Such Euclidean bounce-like configurations are responsible for some tunneling process, say, monopole--antimonopole pair production.
 The information concerning the torus knot invariants is  stored in this Euclidean configuration before the
analytic continuation to the Minkowski space. Could we recognize  the knot invariants upon tunneling in 
the Minkowski space? We plan  to discuss this point in the separate publication.

\end{itemize}

 The authors are grateful to   E. Gorsky, S. Gukov, P. Koroteev, N. Nekrasov, A. Zabrodin and A. Zotov
 for discussions. The work of A.G. and K.B.  was supported in part
by grants RFBR-12-02-00284 and PICS-12-02-91052. The work of K.B. was also supported by the Dynasty fellowship program. We thank the
organizers of  Simons Summer School at Simons Center for Geometry
and Physics where the part of this work has been done for the
hospitality and support. We thank the IPhT at Saclay where the part of this work 
has been done for the
hospitality and support.

\appendix

\section{HOMFLY polynomial and theory on the string worldsheet}
\label{app_homfly}
In this section we briefly remind how to compute the HOMFLY polynomial using the action of the Cherednik algebra on the symmetric polynomials \cite{G11,G12}. Let $x(z), y(z)$ be polynomials describing embedding of an $(m,n)$ (we change 
$(p,q)\rightarrow (n,m)$ in Appendix)  Seifert surface into $\mathbb C^2$:

\begin{equation}
    x^m(z)=y^n(z)\Leftrightarrow (1+u_2z^2+\ldots+u_nz^n)^{\frac mn}=1+v_2z^2+\ldots+v_mz^m.
    \label{Seif}
\end{equation}

Here we can think of $z\in \mathbb C$ as of the worldsheet coordinate of the open topological string with the Seifert surface
in the target space in the spirit of \cite{ooguri}. Let $J_{m/n}$ denote an ideal in $\mathbb C[u_2,\ldots,u_n]$ generated by the coefficients in the Taylor expansion of $(1+u_2z^2+\ldots+u_nz^n)^{\frac mn}$ from $(m+1)$'th to $(m+n-1)$'th. Let us introduce the space $\mathcal M_{m,n}=\mathbb C[u_2,\ldots,u_n]/J_{m/n}$ and differential forms on this space $\Omega^\bullet(\mathcal M_{m,n})$.

{\it Example.} For $T_{2k+1,2}$ knots the construction above gives:
\begin{equation}
    J_{(2k+1)/2}=\langle u_2^{k+1}\rangle.
    \label{J_ex}
\end{equation}

The space $\Omega^\bullet(\mathcal M_{m,n})$ is generated by forms $(1,u_2,\ldots, u_2^{k}, du_2,\ldots,u_2^{k-1}du_2)$.

The HOMFLY polynomial can be represented as a graded trace over the space $\Omega^\bullet(\mathcal M_{m,n})$:

\begin{equation}
    P_{m,n}(a,q)=a^{(n-1)(m-1)}\sum_{i=0}^{n-1}(-a^2)^i\tr(q^K; \Omega^i(\mathcal M_{m,n})).
    \label{homfly}
\end{equation}

Here $K$ is the dilatation operator from (\ref{sl2}). The Dunkl operators acts on $u_i$ as following: we identify $x_i$ in (\ref{def_Dunkl}) with the inverse roots of the $x(z)$ polynomial,

\begin{equation}
    x(z)=\prod_{i=1}^{n}(1-x_iz).
    \label{x_z}
\end{equation}

Then $u_i$ are symmetric polynomial in $x_i$.

{\it Example.} For $T_{2k+1,2}$ knots

\begin{equation}
    u_2=x_1x_2,\qquad du_2=x_1dx_2+x_2dx_1.
    \label{u(x)}
\end{equation}


The other way to say the same thing is the following: consider the polynomials $\mathbb C\left[ x_1,\cdots,x_n \right]$ on which the Dunkl operator (\ref{def_Dunkl}) acts. Let the center of mass of the $n$-particle system be at zero and consider the polynomials $P(x_1,\cdots,x_n)$ which do not depend on the center of mass coordinate:

\begin{equation}
    \sum_{i=1}^n \mathcal D_i P=0.
    \label{COM}
\end{equation}

Consider the polynomials which are annihilated by the Dunkl operator and the ideal $I_{m/n}$ generated by these polynomials. Then the HOMFLY polynomial (\ref{homfly}) can be computed from the action of the dilation operator on the space $L_{m/n}=\mathbb C[x_1,\cdots,x_n]/I_{m/n}$. One grading counts the degree of the polynomial and the other reflects the representation of the permutation group in which acts on the polynomial \cite{G12},

\begin{equation}
    P_{m,n}(a,q)=a^{(m-1)(n-1)}\sum_{i=1}^{n-1}(-a^2)^i \dim_q \Hom_{S_n}\left( \Lambda^i \mathfrak h,L_{m/n} \right),
    \label{homfly1}
\end{equation}
where $\mathfrak h$ is a space spanned on the Dunkl operators, $\mathfrak h=\langle \mathcal D_1, \ldots, \mathcal D_n \rangle$, and $\Lambda^i$ is the $i$-th exterior power.

{\it Example.} Let us once again turn to example of $T_{2k+1,2}$ torus knots. Under the condition $x_1+x_2=0$ the space we are considering becomes the space of the polynomials depending on one variable $\mathbb C[x]$. The Dunkl operators annihilate $x^{2k+1}$, hence

\begin{equation}
    I_{(2k+1)/2}=\langle x^{2k+1}\rangle, \qquad L_{(2k+1)/2}=(1,x,\ldots,x^{2k}).
    \label{I1}
\end{equation}

The permutation group $S_2$ has a symmetric and an anti-symmetric representations, hence the $a$-grading distinguishes odd powers from even ones. 

The expressions (\ref{homfly}, \ref{homfly1}) give the HOMFLY polynomial in the normalization where the skein relation is the following:

\begin{equation}
    aP^+(a,q)-a^{-1}P^-(a,q)=(q^{1/2}-q^{-1/2})P^0(a,q),
    \label{skein}
\end{equation}

where $P^+$ denotes the HOMFLY polynomial for a knot with an ``undercrossing'', $P^-$ is the polynomial for the knot with an ``overcrossing'', and $P^0$ is the polynomial for the knot without that crossing. The HOMFLY polynomial for a $T_{2k+1,2}$ torus knot is:

\begin{equation}
    P_{2k+1,2}=a^2\sum_{i=0}^{k}q^{2i}-\sum_{i=0}^{k-1}q^{2i+1},
    \label{homfly_torus}
\end{equation}

as can be verified from (\ref{homfly}, \ref{homfly1}) using the (\ref{J_ex}, \ref{u(x)},\ref{I1}) expressions.

This form of the polynomial for the $T_{2k+1,2}$ torus knots suggests an interpretation in terms of a modification of the Witten index for some quantum--mechanical system. Indeed, for the polynomials of one variable we can write the Dunkl operators as follows:

\begin{equation}
    \mathcal D=\frac{\partial}{\partial x}+\frac{2k+1}2\frac{(-1)^P-1}{x},
    \label{dunkl_2}
\end{equation}

where $P$ is the parity operator. The polynomials in the factor $\mathbb C[x]/\langle x^{2k+1}\rangle$ can be formally distinguished into ``fermions'' and ``bosons'' which have eigenvalues $-1$ or $1$ under the parity transformation. The operators $(x, \mathcal D, K=x\mathcal D+\mathcal D x)$ form an $sl_2$ algebra (note that this algebra is different from (\ref{sl2})) which can be understood as algebra of supercharges $Q, Q^\dagger, H=QQ^\dagger$ in a supersymmetric quantum mechanics. The ``raising'' and ``lowering'' operators $(x,\mathcal D)$ map between ``bosons'' and ``fermions''. Note that here $H$ is not the Calogero Hamiltonian, but is a Hamiltonian in some auxiliary quantum problem. The HOMFLY polynomial appears to be a one-parametric modification of the Witten index:

\begin{equation}
    W=\sum_{\mathbb C[x]/\langle x^{2k+1}\rangle} (-1)^F q^H\longrightarrow P_{2k+1,2}=-a\sum_{\mathbb C[x]/\langle x^{2k+1}\rangle} (-a)^F q^H.
    \label{witten}
\end{equation}

The ``spectrum'' is bounded by the condition $\mathcal D \psi=0$ which is solved by a constant and a $x^{2k+1}$ monomial. This hints that the HOMFLY polynomial in principle can be considered as some invariant of generic supersymmetric quantum mechanical or quasiexactly solvable system. We hope to discuss this issue elsewhere.


\thebibliography{99}
  \bibitem{bcgk}
      K.~Bulycheva, H.~-Y.~Chen, A.~Gorsky and P.~Koroteev, ``BPS States in Omega Background and Integrability,''
  JHEP {\bf 1210}, 116 (2012)
  [arXiv:1207.0460 [hep-th]].

\bibitem{ito}
  K.~Ito, S.~Kamoshita and S.~Sasaki,
  ``BPS Monopole Equation in Omega-background,''
  JHEP {\bf 1104}, 023 (2011)
  [arXiv:1103.2589 [hep-th]].

\bibitem{hellerman}
  S.~Hellerman, D.~Orlando and S.~Reffert,
  ``BPS States in the Duality Web of the Omega deformation,''
  JHEP {\bf 1306}, 047 (2013)
  [arXiv:1210.7805 [hep-th]].

\bibitem{gukwit}
  S.~Gukov and E.~Witten,
  ``Gauge Theory, Ramification, And The Geometric Langlands Program,'' arXiv:hep-th/0612073.

\bibitem{dgg}
  N.~Drukker, D.~Gaiotto and J.~Gomis,
  ``The Virtue of Defects in 4D Gauge Theories and 2D CFTs,''
  JHEP {\bf 1106}, 025 (2011)
  [arXiv:1003.1112 [hep-th]].
\bibitem{agt} 
  L.~F.~Alday, D.~Gaiotto and Y.~Tachikawa,
  ``Liouville Correlation Functions from Four-dimensional Gauge Theories,''
  Lett.\ Math.\ Phys.\  {\bf 91}, 167 (2010)
  [arXiv:0906.3219 [hep-th]].
\bibitem{agt2} Alday, L. F., Gaiotto, D., Gukov, S., Tachikawa, Y., Verlinde, H. (2010). Loop and surface operators in $\mathcal {N}= 2$ gauge theory and Liouville modular geometry, Journal of High Energy Physics, 2010(1), 1-50, arXiv:0909.0945 [hep-th].
\bibitem{knots1}
  E.~Witten,
  ``Fivebranes and Knots,''
  arXiv:1101.3216 [hep-th].

\bibitem{knots2}
  D.~Gaiotto and E.~Witten,
  ``Knot Invariants from Four-Dimensional Gauge Theory,''
  Adv.\ Theor.\ Math.\ Phys.\  {\bf 16}, no. 3, 935 (2012)
  [arXiv:1106.4789 [hep-th]].
\bibitem{torus1}
  A.~Brini, B.~Eynard and M.~Marino,
  ``Torus knots and mirror symmetry,''
  Annales Henri Poincare {\bf 13}, 1873 (2012)
  [arXiv:1105.2012 [hep-th]].
\bibitem{torus2}
  M.\,Aganagic, S.\,Shakirov,  ``Knot Homology from Refined Chern-Simons Theory,'' arXiv:1105.5117 [hep-th].
 \bibitem{torus3}
 P.~Dunin-Barkowski, A.~Mironov, A.~Morozov, A.~Sleptsov and A.~Smirnov,
  ``Superpolynomials for toric knots from evolution induced by cut-and-join operators,''
  JHEP {\bf 03}, 021 (2013)
   [arXiv:1106.4305 [hep-th]].

\bibitem{gukov2}
  T.~Dimofte, D.~Gaiotto and S.~Gukov,
  ``Gauge Theories Labelled by Three-Manifolds,''
  arXiv:1108.4389 [hep-th].
\bibitem{yamazaki} 
  Y.~Terashima and M.~Yamazaki,
  ``Semiclassical Analysis of the 3d/3d Relation,''
  Phys.\ Rev.\ D {\bf 88}, 026011 (2013)
  [arXiv:1106.3066 [hep-th]].

\bibitem{G11}E.\,Gorsky, ``Arc Spaces and DAHA representations,'' Selecta Mathematica 19 (2013), no. 1, 125-140, arXiv:1110.1674[math.AG].
\bibitem{G12}E.\,Gorsky, A.\,Oblomkov, J.\,Rasmussen, V.\,Shende, ``Torus Knots and the Rational DAHA,'' arXiv:1207.4523[math.RT].
\bibitem{cardy}J.\,Cardy, ``Calogero-Sutherland model and bulk boundary correlations in conformal field theory'', Phys.Lett. B582 (2004) 121-126, arXiv:hep-th/0310291.
\bibitem{gk}
    D.~Gaiotto and P.~Koroteev, ``On Three Dimensional Quiver Gauge Theories and Integrability,''
  JHEP {\bf 1305}, 126 (2013)
  [arXiv:1304.0779 [hep-th]].

\bibitem{Givental} A. Givental and B. Kim, "Quantum cohomology of flag manifolds and Toda
lattices",     Commun. Math. Phys. 168 (1995) 609--642,
arXiv:hep-th/9312096.
\bibitem{MTV1} E. Mukhin, V. Tarasov, A. Varchenko, "KZ Characteristic Variety as the Zero Set of
Classical Calogero-Moser Hamiltonians",     SIGMA 8 (2012), 072,
arXiv:1201.3990 [math.QA].
\bibitem{GZZ} A.\,Gorsky, A.\,Zabrodin, A.\,Zotov, ``Spectrum of Quantum Transfer Matrices via Classical Many-body Systems'' , JHEP 1401 (2014).
\bibitem{krichever} I.\,Krichever, O.\,Lipan, P.\,Wiegmann, A.\,Zabrodin ``Quantum integrable models and discrete classical Hirota equations,'' Comm. Math. Phys. 188(2), (1997) 267-304.
\bibitem{matsuo} A.\,Matsuo, ``KZ Type Equations and Zonal Spherical Functions,'' Preprint of RIMS, Kyoto, 1991.
\bibitem{cherednik} I.\,Cherednik, ``A unification of Knizhnik-Zamolodchikov and Dunkl operators via affine Hecke algebras,'' Inventiones Mathematicae, 106(1) (1991) 411-431.
 \bibitem{veselov} A.\,Veselov, ``Calogero quantum problem, Knizhnik-Zamolodchikov equation, and Huygens principle,'' Theor.Math.Phys. 98, i.3 (1994) 368-376.
\bibitem{etingof} D.\,Calaque, B.\,Enriquez, P.\,Etingof, ``Universal KZB equations I: the elliptic case,'' arXiv:math/0702670 [math.QA].
    \\
    P.\,Etingof, E.\,Gorsky, I.\,Losev, ``Representations of Rational Cherednik algebras with minimal support and torus knots'',  arXiv:1304.3412 [math.RT].
\bibitem{harnad}
M. R. Adams, J. Harnad, J. Hurtubise, ``Dual moment maps into loop algebras,'' Lett. Math. Phys., Vol. 20,
Num. 4, 299-308 (1990)
\\
 J. Harnad, `` Dual Isomonodromic Deformations and Moment Maps to Loop Algebras,'' Commun. Math. Phys., Vol. 166,
Num. 2, 337-365 (1994), arXiv:hep-th/9301076.
\\
 J. Harnad, ``Quantum
isomonodromic deformations and the Knizhnik--Zamolodchikov
equations'', CRM Proc.  Lecture Notes 155-161 (Amer. Math.  Soc. ,
Providence, RI, (1996), arXiv:hep-th/9406078.

\bibitem{MTV3} E.\,Mukhin, V.\,Tarasov, A.\,Varchenko, ``Bispectral and $(gl_N, gl_M)$ Dualities, Discrete Versus Differential,'' arXiv:math/0605172 [math.QA].


\bibitem{Nekrasov} N.\,A.\,Nekrasov, ``Seiberg-Witten Prepotential from Instanton Counting,'' Adv. Theor. Math. Phys. 7 (2004) 831, arXiv:hep-th/0206161.
\bibitem{NekrasovOkounkov} N.\, Nekrasov, A.\,Okounkov, ``Seiberg-Witten Theory and Random Partitions,'' arXiv:hep-th/0306238.
\bibitem{ShifmanYung} M.\,Shifman, A.\,Yung, ``Supersymmetric Solitons and How They Help Us Understand Non-Abelian Gauge Theories,'' Rev. Mod. Phys. 79, 1139 (2007), hep-th/0703267.
\bibitem{ShifmanYung1}M.\,Shifman, A.\,Yung, ``Domain Walls and Flux Tubes in N=2 SQCD: D-Brane Prototypes,'' Phys.Rev.D67:125007, 2003, arXiv:hep-th/0212293.
\bibitem{ShifmanYung2}M.\,Shifman, A.\,Yung, ``Localization of Non-Abelian Gauge Fields on Domain Walls at Weak Coupling (D-brane Prototypes)'', Phys.Rev.D70:025013, 2004, arXiv:hep-th/0312257.

\bibitem{gs} A.\,Gorsky, and M.\,Shifman, ``More on the tensorial central charges in N=1
    supersymmetric gauge theories (BPS wall junctions and strings)'', Phys. Rev. D61 (2000) 085001, {arXiv:hep-th/9909015}.
 \bibitem{ns1}
      N.~A.~Nekrasov and S.~L.~Shatashvili, ``Supersymmetric vacua and Bethe ansatz,''
  Nucl.\ Phys.\ Proc.\ Suppl.\  {\bf 192-193}, 91 (2009)
  [arXiv:0901.4744 [hep-th]].

  \bibitem{ns2}
      N.~A.~Nekrasov and S.~L.~Shatashvili, ``Quantization of Integrable Systems and Four Dimensional Gauge Theories,''
  arXiv:0908.4052 [hep-th].
 \bibitem{nw}
      N.~Nekrasov and E.~Witten, ``The Omega Deformation, Branes, Integrability, and Liouville Theory,''
  JHEP {\bf 1009}, 092 (2010)
  [arXiv:1002.0888 [hep-th]].
\bibitem{nrs}
    N.~Nekrasov, A.~Rosly and S.~Shatashvili, ``Darboux coordinates, Yang-Yang functional, and gauge theory,''
  Nucl.\ Phys.\ Proc.\ Suppl.\  {\bf 216}, 69 (2011)
  [arXiv:1103.3919 [hep-th]].
\bibitem{gornek}
    A.~Gorsky and N.~Nekrasov, ``Relativistic Calogero-Moser model as gauged WZW theory,''
  Nucl.\ Phys.\ B {\bf 436}, 582 (1995)
  [hep-th/9401017].

\bibitem{gw1}
    D.~Gaiotto and E.~Witten, ``S-Duality of Boundary Conditions In N=4 Super Yang-Mills Theory,''
  Adv.\ Theor.\ Math.\ Phys.\  {\bf 13} (2009)
  [arXiv:0807.3720 [hep-th]].
  \bibitem{gw2}
      D.~Gaiotto and E.~Witten, ``Supersymmetric Boundary Conditions in N=4 Super Yang-Mills Theory,''
  J.\ Statist.\ Phys.\  {\bf 135}, 789 (2009)
  [arXiv:0804.2902 [hep-th]].

\bibitem{BPZ}Belavin, Alexander A., Alexander M. Polyakov, and Alexander B. Zamolodchikov. ``Infinite conformal symmetry in two-dimensional quantum field theory.'' Nuclear Physics B 241.2 (1984): 333-380.


\bibitem{KZ}Knizhnik, V. G., and A. B. Zamolodchikov. ``Current algebra and Wess-Zumino model in two dimensions.'' Nuclear Physics B 247.1 (1984): 83-103.

\bibitem{FZ} Zamolodchikov, A. B., and V. A. Fateev. ``Operator algebra and correlation functions in the two-dimensional SU (2) x SU (2) chiral Wess-Zumino model.'' Sov. J. Nucl. Phys.(Engl. Transl.);(United States) 43.4 (1986).


\bibitem{hanany_witten} A.\,Hanany and E.\,Witten, ``Type IIB superstrings, BPS monopoles, and three-dimensional
    gauge dynamics,'' Nucl.Phys. B492, 152 (1997), hep-th/9611230.

\bibitem{gorneg} E.\,Gorsky, A.\,Negut, ``Refined knot invariants and Hilbert schemes,'' arXiv:1304.3328 [math.RT].
\bibitem{holland} 
 T.~Dimofte, S.~Gukov and L.~Hollands,
  ``Vortex Counting and Lagrangian 3-manifolds,''
  Lett.\ Math.\ Phys.\  {\bf 98}, 225 (2011)
  [arXiv:1006.0977 [hep-th]].
\bibitem{ooguri} 
  H.~Ooguri and C.~Vafa,
  ``Knot invariants and topological strings,''
  Nucl.\ Phys.\ B {\bf 577}, 419 (2000)
  [hep-th/9912123].
\end{document}